\documentclass[12pt]{article}

\makeatletter
\@addtoreset{equation}{section}
\makeatother

\newcommand{\be}{\begin{equation}}
\newcommand{\ee}{\end{equation}}
\newcommand{\bee}{\begin{eqnarray}}
\newcommand{\beee}{\begin{array}}
\newcommand{\eee}{\end{eqnarray}}
\newcommand{\eeee}{\end{array}}

\newcommand{\vac}{|0\rangle\langle 0|}%
\newcommand{\z}{\,,\qquad}

\newcommand{\ga}{\alpha}
\newcommand{\gb}{\beta}
\newcommand{\gga}{\gamma}

\newcommand{\M}{{\cal M}}
\newcommand{\ls}{\!\!\!\!\!\!}

\newcommand{\gd}{\delta}

\newcommand{\gep}{\epsilon}

\newcommand{\gs}{\sigma}

\newcommand{\nn}{\nonumber}
\newcommand{\half}{\frac{1}{2}}

\newcommand{\p}{\partial}

\newcommand{\f}{\frac}

\begin{document}
\begin{flushright}
\vspace{1mm}
FIAN/TD/06--02\\

{April 2002}\\
\end{flushright}

\vspace{1cm}

\begin{center}
{\Large {\bf Higher Spin Conserved Currents\\ in  $Sp(2M)$
Symmetric  Spacetime}}

\vspace{1cm}
{\bf  M.A.~Vasiliev }\footnote{e-mail: vasiliev@lpi.
ru}  \\
\vspace{1cm}

I.E.Tamm Department of Theoretical Physics, Lebedev Physical
Institute,\\
Leninsky prospect 53, 119991, Moscow, Russia
\vspace{1.5cm}
\end{center}

\vspace{1cm}

\begin{abstract}
Infinite set of higher spin conserved charges is found
for the $sp(2M)$ symmetric dynamical systems
in $\f{1}{2} M (M+1)$-dimensional generalized spacetime $\M_M$.
Since the dynamics in $\M_M$ is equivalent to the
conformal dynamics of infinite towers of fields in $d$-dimensional
Minkowski spacetime with
$d=3,4,6,10,\ldots$ for $M=2,4,8,16,  \ldots$,
respectively, the constructed currents in $\M_M$ generate infinite
towers of (mostly new) higher spin
conformal currents in Minkowski spacetime.
The charges have a form of integrals of
$M$-forms which are bilinear in the field variables and are
closed as a consequence of the field equations. Conservation
implies independence of a value of charge of a local variation of a
 $M$-dimensional integration surface $\Sigma \subset \M_M$
analogous to Cauchy  surface in the usual spacetime.
The scalar conserved charge provides an  invariant bilinear
form on the space of solutions of the field equations that gives rise to
a positive definite norm on the space of quantum states.

\end{abstract}

\section{Introduction}\label{Generalities}

The idea that symplectic superalgebras
$osp(1,2^p)$ and their subalgebras and contractions
are important for understanding dualities and
$M$-theory is attractive as these algebras contain
supergravity algebras in diverse dimensions in a natural way
\cite{AF,HP,T,B1,S,B2,T2,GM,G,FP,FL,BP,AFLV,AFL,FL}.
On the other hand, symplectic superalgebras
were argued \cite{F,Vfer,BHS,Mar} to play a key role in the theory of
massless higher spin fields.
In particular, Fronsdal emphasized \cite{F} that the set of
$4d$ massless
fields of all spins should exhibit $sp(8)$ symmetry and argued that some
formulation of their dynamics in the generalized spacetime with matrix
coordinates $X^{\ga\gb}=X^{\gb\ga}$ ($\ga, \gb \ldots = 1,\ldots , 4$)
must exist.

Spacetimes with symmetric real matrix coordinates
provide a natural realization of the symplectic algebras \cite{H,F}.
Relevant constructions of extended spacetimes
 were discussed by many authors in different contexts
\cite{Curt,C,ES,G1,RS,DN,BL,BLS,CAIP,Ca,GGHT,WN,ZL,P,BAIL}.
{}From $M$-theory perspective,
 relevance of generalized spacetimes with matrix
coordinates $X^{\ga\gb}=X^{\gb\ga}$ ($\ga, \gb \ldots = 1,\ldots , 2^p$)
is due to observation that once  $\ga,\gb \ldots$ are
interpreted as spacetime spinor indices the matrix coordinates
$X^{\ga\gb}$ provide a set of antisymmetric tensor
``central charge coordinates"  $x^{n_1 \ldots n_k}$
dual to ``central charges" $Z^{n_1 \ldots n_k}$  and spacetime
momenta resulting from the decomposition of the anticommutator of
supercharges $\{Q_\ga , Q_\gb \}$ into irreducible Lorentz tensors
\be
\label{CC}
\{Q_\ga , Q_\gb \} = \sum_{k\in S} \gga_{n_1 \ldots n_k}{}_{\ga\gb}
Z^{n_1 \ldots n_k}\,.
\ee
Here summation is over those values
$k \in S$ that totally antisymmetrized products of
 $\gga$-matrices  $\gga_{n_1 \ldots n_k}{}_{\ga\gb}$ are symmetric
in the spinor indices $\ga,\gb$. Since ``central  charges"
$Z^{n_1 \ldots n_p}$ characterize
branes in superstring theory, a unified treatment
of brane dualities requires uniform description of all
``central charges". This is achieved by
introducing ``central charge coordinates"
\cite{Curt,RS,BL,BLS,CAIP,Ca,GGHT,WN,ZL,BAIL} which together with the
usual coordinates are equivalent to the coordinates $X^{\ga\gb}$.
A remarkable observation that
supergravity models may result from spontaneous breakdown
of symplectic symmetries was made recently in \cite{BW,WEST}
where it was shown that the equations of
motion of $11d$ supergravity imply conservation conditions for some
$osp(1,64)$ currents while the theory as a whole provides a nonlinear
realization of $osp(1,64)$.

The difference between the role of $sp(2^p)$
and its superextensions in the $M$-theory setup  and in
higher spin gauge theory
(see \cite{rev,Gol} for a review  and more references on higher spin
gauge theories) is that $M$-symmetries
$sp(2^p)$ are broken (nonlinearly realized) while in the higher spin
gauge theory  they act linearly and locally on fields
as was recently shown in \cite{BHS}. Higher symplectic symmetries
in higher spin gauge theories (e.g., $sp(8)$ is $4d$ higher spin
theory) mix massless fields of all spins. As all massless higher spin
fields are gauge fields this implies that a linearly realized higher
symplectic symmetry is only possible in an
invariant phase exhibiting infinite-dimensional
higher spin symmetry that forms an infinite-dimensional extension of the
symplectic symmetry.
Since higher spin modes in superstring are massive, this explains
why higher symplectic symmetries can only show up via nonlinear
realization in the low-energy supergravity models.

The idea that higher spin gauge theory is a natural candidate for a
most symmetric phase of the theory of fundamental interactions presently
identified with superstring theory and $M$-theory provided originally the
main motivation for its investigation \cite{Fort1}.
A peculiar feature of the higher spin gauge theory is that gauge invariant
higher spin interactions require nonzero cosmological constant
\cite{FV1}. This most symmetric phase differs from the phases of superstring  
with known spectra like in the flat space \cite{GSW} or pp wave
background \cite{Met}. The (so far) explicitly
solvable phases of superstring therefore require
higher spin symmetries to be broken. It is tempting to speculate
that $M-$theory and superstring theory may
result from a higher spin gauge theory via spontaneous breakdown
of a higher spin symmetry that contains symplectic symmetry
linking together
higher spin and lower spin fields. Some ideas on a possible
connection between higher spin theory and string theory
in the context of $AdS/CFT$ correspondence were recently discussed in
\cite{Su,Witten,SS,BHS,Mih,Ts}.
Once a nonlinear symplectic supersymmetry in $M$-theory is
indeed a manifestation of a higher spin
symmetric phase in which it is unbroken, this implies that
branes are built of (vev's of) higher spin gauge fields \cite{BHS}.
In other words, higher spin gauge theories are expected to provide a
microscopic description of branes.

For geometric realization of
symplectic symmetries, dynamics has to be reformulated  in terms of
the generalized  $\half M (M+1)-$dimensional spacetime $\M_M$
with real symmetric matrix coordinates $X^{\ga\gb}=X^{\gb\ga}$
$(\ga,\gb = 1,\ldots M$) \cite{H,F,GGHT,BHS,Mar}, in which
infinitesimal $Sp(2M)$ transformations
are realized by the vector fields \cite{BHS}
\be
\label{PS}
 P_{\ga\gb} =-i \f{\p}{\p X^{\ga\gb}}\,,
\ee
\be
L_\ga{}^\gb = 2i
 X^{\gb\gga} \f{\p}{\p X^{\ga \gga}}\,,
\ee
\bee
\label{PK}
K^{\ga\gb}=  -i X^{\ga\gga}
X^{\gb{\eta}}\f{\p}{\p X^{\gga{\eta}}}\,.
\eee
The (nonzero) $sp(2M)$ commutation relations are
\be
\label{comLL}
[L_\ga{}^\gb \,, L_\gga{}^\gd] = i\left( \gd^\gd_\ga L_\gga{}^\gb -
\gd^\gb_\gga L_\ga{}^\gd \right )\,,
\ee
\be
\label{comL}
[L_\ga{}^\gb \,, P_{\gga\gd}] = -i \left (\gd^\gb_\gga P_{\ga\gd}
+\gd^\gb_\gd P_{\ga\gga}\right )\,,\qquad
[L_\ga{}^\gb \,, K^{\gga\gd}] =i \left (\gd^\gga_\ga K^{\gb\gd}+
\gd_\ga^\gd K^{\gb\gga}\right )\,,\qquad
\ee
\be
\label{comPK}
[P_{\ga\gb} \,, K^{\gga\gd}] = \f{i}{4}
\Big (\gd_\gb^\gga L_\ga{}^\gd + \gd_\ga^\gga L_\gb{}^\gd +
\gd_\ga^\gd L_\gb{}^\gga + \gd_\gb^\gd L_\ga{}^\gga \Big ) \,.
\ee

Here $P_{\ga\gb}$ and $K^{\ga\gb}$ are generators of the
generalized translations and special conformal transformations.
The $gl_M$ algebra spanned by $L_\ga{}^\gb $
decomposes into the central subalgebra associated with the
generalized dilatation generator
\be
D= L_\ga{}^\ga \nn
\ee
 and the  $sl_M$ generalized  Lorentz generators
\be
l_\ga{}^\gb = L_\ga{}^\gb -\f{1}{M}\delta^\gb_\ga  D\,.\nn
\ee

{\it A priori}, it is not obvious how to formulate consistent $sp(2M)$
invariant dynamical equations compatible with the standard principles of
quantum field theory such as unitarity and causality.
Some proposals were made, e.g, in \cite{C,Ca,GGHT}.
An obvious problem is that it is hard to write down
an analog of the Klein-Gordon equation free from the ghost problem
because  each``central charge momentum" $Z_{n_1 \ldots n_k}$
induces both positive an negative contributions
to any Lorentz invariant norm.

On the other hand, the ``unfolded formulation"
of the free conformal higher spin fields in $d=3$ \cite{3d} and $d=4$
\cite{BHS} allowed us to derive a form
of $sp(2M)$ invariant equations in $\M_M$ \cite{BHS} equivalent to
the usual higher spin equations compatible with unitarity.
As a result of this reformulation, massless fields of all integer
spins in four dimensions are described by a single scalar field $c(X)$ in
$\M_M$.  All half-integer spins are described by
a single svector field $c_\gb(X)$.
(We use the name ``svector" (symplectic vector) to distinguish $c_\gb(X)$
from  vectors of the usual Lorentz algebra $o(d-1,1)$.  Note that
svector fields
obey the Fermi statistics \cite{Mar}).  The $sp(2M)$ invariant equations of
motion found in \cite{BHS} which encode $4d$ massless equations for all spins
read
\be
\label{oscal} \Big ( \f{\p^2}{\p X^{\ga\gb} \p X^{\gga\gd}} - \f{\p^2}{\p
X^{\ga\gga} \p X^{\gb\gd}}\Big ) c(X) =0
\ee for a scalar field $b(X)$ and
\be \label{ofer} \f{\p}{\p X^{\ga\gb}} c_\gga(X) -
\f{\p}{\p X^{\ga\gga}} c_\gb(X) =0
\ee
for a svector field $c_\gb(X)$.  Note that the same equations were argued in
\cite{BHS,Mar} to make sense for any even number $M$ of values taken by svector
indices $\ga, \gb = 1,\ldots M$.
For $M=2$, because antisymmetrization of any two-component indices $\ga$ and
$\gb$ is equivalent to their contraction with the $2\times 2 $ symplectic form
$\gep^{\ga\gb}$, (\ref{oscal}) and (\ref{ofer}) coincide with the $3d$ massless
Klein-Gordon and Dirac equations, respectively.

Properties of the equations (\ref{oscal}) and (\ref{ofer}) were
analyzed in detail in \cite{Mar} where
the dynamics in $\M_M$
described by the equations (\ref{oscal}) and (\ref{ofer})
was shown to be consistent
with the principles of relativistic quantum field theory including
unitarity and microcausality. The most important difference as compared
to the usual picture is that, because the system of equations
(\ref{oscal}) and (\ref{ofer}) is overdetermined, true local phenomena
occur in a smaller space $\gs$ called local Cauchy surface
in \cite{Mar}
and identified with the space slice of Minkowski spacetime.
The dependence  along all ``time-like" directions turns out
to be fixed at once in terms of appropriate initial data.
The full set of ``initial data" that fixes a solution of the field equations
(\ref{oscal}) and (\ref{ofer})  is provided by two functions on a
$M$-dimensional ``local Cauchy bundle" $E$ having local Cauchy surface
$\gs$ as its base manifold. From the point of view of usual
Minkowski geometry the fiber space of $E$ parametrizes spin degrees of
freedom of the fields living in the Minkowski spacetime
$\gs\times R$. Note that it is appropriate to describe local phenomena
in the Minkowski spacetime $\gs\times R$ in terms of the local Cauchy
bundle $E$ rather than in terms of some $M$-dimensional surface in $\M_M$.
The difference is that $E$ is a limit of
some surface $\gs\times \tau$ with the size of $\tau$ tending to zero.
The resulting limiting description in terms of $E$ becomes local in terms of
$\gs$ \cite{Mar}.

The formulations of $Sp(2M)$ invariant systems in terms of the generalized
space-time $\M_M$ and usual spacetime are equivalent and
complementary. The description in terms of $\M_M$ provides clear geometric
origin for the $Sp(2M)$ generalized conformal symmetry.  In particular it
provides geometric interpretation of the electromagnetic duality
transformations as particular generalized Lorentz transformations \cite{Mar}.
The description in terms of the Minkowski spacetime admits standard
Cauchy problem but makes some of the symmetries not manifest, namely
those which do not leave $E$ invariant. The description in terms of
different local Cauchy surfaces $\gs$ and associated local Cauchy bundles
$E$ are equivalent being related by some $Sp(2M)$ transform.
One can say that although the dynamics is formulated in $\M_M$ in
explicitly $sp(2M)$ invariant manner, the generalized spacetime $\M_M$
is visualized by virtue of local phenomena as some $d-1 \leq M$
dimensional space $\gs$ times a $M-d+1$ dimensional
fiber associated with spin degrees of freedom.
This picture has striking similarities with the brane picture.
To work out a full-scale correspondence it is necessary to develop full
nonlinear theory of higher spin gauge fields in $\M_M$.

In this paper we
make a modest step in this direction by constructing conserved charges
built of the fields $c(X)$ and $c_\ga (X)$ in $\M_M$. The
constructed currents are in
the one-to-one correspondence with the set of generalized
higher spin conformal symmetries found in \cite{BHS} which contain
$ OSp(L,2M)$ symmetries as a subgroup. Due to
specificity of the Cauchy problem in $\M_M$, the corresponding integrals
of motion have a form of integrals of certain on-mass-shell
closed $M-$forms $\Omega$. Being independent
of a choice of a $M-$dimensional integration surface
they provide ``integrals of motion" that characterize a particular
solution of the field equations. In view
of the results of \cite{BW,WEST} the conservation conditions
of the currents associated with the symplectic superalgebras
may be related to the equations of motion in $M$-theory
in a higher spin Higgs phase.

The rest of the paper is organized as follows. In section
\ref{General Structure of Currents} a set of integrals of
motions is built provided that a generalized stress tensor
satisfying appropriate conservation conditions exists. In section
\ref{Generalized Stress Tensors} the generalized stress tensor is built
in terms of bilinears of the dynamical fields in $\M_M$. In section
\ref{Fourier Transform and Invariant Norm} the expressions
for the conserved charges are evaluated in terms of Fourier transformed
field variables and it is
shown that they reproduce the expressions for the generators of
higher spin transformations derived previously in \cite{BHS}.
Section \ref{Conclusion} contains brief conclusions.

\section{General Structure of Currents}
\label{General Structure of Currents}

The fact \cite{Mar} that dynamical degrees of freedom associated with the
equations (\ref{oscal}) and (\ref{ofer}) live on a $M$-dimensional
subsurface $S\subset \M_M$ or its limiting $M$-dimensional local
Cauchy bundle $E$ suggests that integrals of motion associated with
these equations have to be built in terms of some $M$-forms
$\Omega (\eta)$ which are closed
\be
d\Omega (\eta) =0\,,\qquad d=dX^{\ga\gb} \f{\p}{\p X^{\ga\gb}}\,
\ee
as a consequence of the field equations (\ref{oscal}) and (\ref{ofer}).
Here $\eta$ are some  parameters associated with
different closed $M$-forms and charges
\be
\label{charge}
Q(\eta) = \int_S \Omega (\eta)\,.
\ee
The charges $Q(\eta) $ are independent of local variations of
$S$ and, therefore, provide a set of integrals of motion associated
with a particular solution of the field equations. In other words,
the charges $Q$ are conserved. They generate
symmetries by Poisson brackets or commutators upon quantization.
(For the quantization rules for the fields $c(X)$ and $c_\ga (X)$ see
\cite{Mar} and section  \ref{Fourier Transform and Invariant Norm}).
The modules $\eta$ are then interpreted as the symmetry
parameters associated with the generators $Q(\eta )$. Note that
addition to $\Omega (\eta )$ an
exact form does not affect charges. This ambiguity
characterizes possible ``improvements".

Note that the dimension of a Cauchy surface in the
$d$-dimensional Minkowski spacetime
is $d-1$. The closed forms $\Omega$ associated with integrals of motion
are of degree $d-1$. The conserved currents $J$ are vector fields
dual to $\Omega$. In the generalized
spacetime $\M_M$ it is natural to formulate the problem in terms of
closed $M-$forms rather than conserved dual polyvectors of rank $\half M(M-1)$.
In the sequel  on-mass-shell closed forms $\Omega$ are called conserved.

We proceed in two steps by analogy with the case of usual conformal
higher spin currents considered in \cite{KVZ}. Firstly,
 we observe that, once there is a totally symmetric multisvector
$T_{\ga_1 \ldots \ga_{n}}$, $n\geq M$ satisfying certain generalized
conservation conditions,
%analogous to the conservation of traceless totally
%symmetric conformal stress tensor and its higher spin analogues,
the  $M$-form
\be
\label{clo} \Omega (\eta ) = \epsilon_{\gamma_1 \ldots \gamma_M}
dX^{\gamma_1 \ga_1}\wedge \ldots \wedge dX^{\gamma_M \ga_M} \eta_{\gb_1 \ldots
\gb_t}{}^{\ga_{M+1} \ldots \ga_{M+s}} X^{\ga_{M+s+1} \gb_1}\ldots X^{\ga_{M+s+t}
\gb_t} T_{\ga_1 \ldots \ga_{M+s+t}}\,
\ee
is closed.
Here $\epsilon_{\gamma_1 \ldots \gamma_M} $ is the totally antisymmetric symbol.
It is introduced to get rid of the set of totally antisymmetric indices from the
parameter $\eta_{\gb_1 \ldots \gb_t}{}^{\ga_{1} \ldots \ga_{s}}$ being an
arbitrary totally symmetric $X-$independent multisvector in lower and upper
indices. At the second stage we will explicitly construct the
generalized stress tensors in terms of bilinears in $c(X)$ and $c_\ga (X)$.

The generalized conservation condition can be written in
the following three equivalent forms
\be
\label{cond}
\f{\p}{\p X^{\gamma\eta}}T_{\ga\gb \ga_3 \ldots \ga_n}
-\f{\p}{\p X^{\gamma\ga}}T_{\eta\gb \ga_3 \ldots \ga_n}
-\f{\p}{\p X^{\gb\eta}}T_{\ga\gamma \ga_3 \ldots \ga_n}
+\f{\p}{\p X^{\gb\ga}}T_{\gamma\eta \ga_3 \ldots \ga_n} =0\,,
\ee
or
\be
\label{cond1}
\lambda^{ \ga_1 \ldots \ga_n\,,\gb_1 \gb_2}
\f{\p}{\p X^{\gb_1 \gb_2}}T_{ \ga_1 \ldots \ga_n}
= 0
\ee
for any $\lambda^{ \ga_1 \ldots \ga_n\,,\gb_1 \gb_2}$ having the
symmetry properties of two-row Young scheme with two cells in the
second row (i.e., $\lambda^{ \ga_1 \ldots \ga_n\,,\gb_1 \gb_2}$ is
separately symmetric in the indices $\ga_i$ and $\gb_j$ and
symmetrization of any $n+1$ indices
gives zero), or
\be
\label{repr}
\f{\p}{\p X^{\gamma\eta}}T_{\ga_1 \ldots \ga_n} =
U_{\ga_1 \ldots \ga_n\gamma , \eta } +
U_{\ga_1 \ldots \ga_n\eta , \gamma } \,,
\ee
where $U_{\ga_1 \ldots \ga_{n+1},\eta }$ is some
multisvector totally symmetric in the
$n+1$ indices $\ga$. The conditions (\ref{cond})-(\ref{repr})
are equivalent to each other because the derivative
$\f{\p}{\p X^{\gamma\eta}}T_{\ga_1 \ldots \ga_n} $ decomposes into three
irreducible parts associated with two-row Young schemes with zero, one and
two cells in the second row
\be
% 1 stroka --2
\begin{picture}(450,35)
\put(11,24){$2$}
{\linethickness{.500mm}
\put(00,20){\line(1,0){20}}%
%\put(00,30){\line(1,0){20}}%
\put(00,10){\line(1,0){20}}%
\put(00,10){\line(0,1){10}}%
\put(10,10.0){\line(0,1){10}} \put(20,10.0){\line(0,1){10}}
\put(23,12){$\bigotimes$}
}
\put(34,00)
{
%1  stroki:n
\begin{picture}(75,35)
\put(30,24){$n$}
{\linethickness{.500mm}
\put(00,10){\line(1,0){70}}%
\put(00,20){\line(1,0){70}}%
\put(00,10){\line(0,1){10}}%
\put(10,10.0){\line(0,1){10}} \put(20,10.0){\line(0,1){10}}
\put(30,10.0){\line(0,1){10}} \put(40,10.0){\line(0,1){10}}
\put(50,10.0){\line(0,1){10}} \put(60,10.0){\line(0,1){10}}
\put(70,10.0){\line(0,1){10}}%\put(80,10.0){\line(0,1){10}}
\put(73,12){$=$}
}
%\put(30,04){$s_2$}
\end{picture}}
\put(120,00)
{
\begin{picture}(95,35)
\put(30,24){$n+2$}
{\linethickness{.500mm}
\put(00,10){\line(1,0){90}}%
\put(00,20){\line(1,0){90}}%
\put(00,10){\line(0,1){10}}%
\put(10,10.0){\line(0,1){10}} \put(20,10.0){\line(0,1){10}}
\put(30,10.0){\line(0,1){10}} \put(40,10.0){\line(0,1){10}}
\put(50,10.0){\line(0,1){10}} \put(60,10.0){\line(0,1){10}}
\put(70,10.0){\line(0,1){10}} \put(80,10.0){\line(0,1){10}}
\put(90,10.0){\line(0,1){10}}%\put(100,10.0){\line(0,1){10}}
\put(93,12){$\bigoplus$}
}
\end{picture}}
\put(324,00)
{
% 2 stroki:n,2
\begin{picture}(70,35)
\put(30,24){$n$}
{\linethickness{.500mm}
\put(00,10){\line(1,0){70}}%
\put(00,20){\line(1,0){70}}%
\put(00,00){\line(1,0){20}}%
\put(00,00){\line(0,1){20}}%
\put(10,00.0){\line(0,1){20}} \put(20,00.0){\line(0,1){20}}
\put(30,10.0){\line(0,1){10}} \put(40,10.0){\line(0,1){10}}
\put(50,10.0){\line(0,1){10}} \put(60,10.0){\line(0,1){10}}
\put(70,10.0){\line(0,1){10}}%\put(80,10.0){\line(0,1){10}}
}
\end{picture}}
\put(227,00)
{
% 2 stroki:n,1
\begin{picture}(80,35)
\put(30,24){$n+1$}
{\linethickness{.500mm}
\put(00,10){\line(1,0){80}}%
\put(00,20){\line(1,0){80}}%
\put(00,00){\line(1,0){10}}%
\put(00,00){\line(0,1){20}}%
\put(10,00.0){\line(0,1){20}}
\put(20,10.0){\line(0,1){10}}
\put(30,10.0){\line(0,1){10}}
\put(40,10.0){\line(0,1){10}}
\put(50,10.0){\line(0,1){10}} \put(60,10.0){\line(0,1){10}}
\put(70,10.0){\line(0,1){10}} \put(80,10.0){\line(0,1){10}}
\put(83,12){$\bigoplus$}
}
\end{picture}}
\end{picture}
%YOUNG
\ee
Each of the  conditions (\ref{cond}), (\ref{cond1}) and (\ref{repr})
implies that the part described by the Young scheme with two cells in
the second row vanishes.

To prove that the $M-$form (\ref{clo}) is closed one observes that,
by virtue of (\ref{repr}), the part of $d\Omega$ due to
differentiation of $T$ vanishes because it is proportional
to $dX^{\gamma_1 \ga_1}\wedge \ldots \wedge dX^{\gamma_{M+1} \ga_{M+1}}
U_{\ga_1 \ldots \ga_{M+1}\ldots\ga_{n+1} , \eta }$ which expression
contains total antisymmetrization over $M+1$ svector indices
$\gamma_i$ taking only
$M$ values. The part due to differentiation of the
factors of $X^{\ga\gb}$ gives rise to the product of $M+1$
differentials $dX^{\ga\gamma}$, each having one index contracted with
the totally symmetric multisvector
$T_{\ga_1 \ldots \ga_{M+s+t}}$.
It  therefore also vanishes because of total antisymmetrization
of the rest $M+1$ svector indices in the anticommuting differentials.

The multisvector $T_{\ga_1 \ldots \ga_n}$ is a generalization
of the spin 1 current, spin 3/2 supercurrent, spin 2
stress tensor and their higher spin extensions \cite{Ans,KVZ}
in the conformal field theory in Minkowski spacetime.
It is notable that the set of the symmetry parameters
$\eta_{\gb_1 \ldots \gb_t}{}^{\ga_{1} \ldots \ga_{s}} $ is in the
one-to-one correspondence with the set of parameters of the
generalized higher spin conformal symmetries found in \cite{BHS} in
the form
\be
\label{eta}
\eta(a,b) = \sum_{n,m=0}^\infty
\eta_{\gb_1 \ldots \gb_t}{}^{\ga_{1} \ldots \ga_{s}}
a_{\ga_1} \ldots a_{\ga_s} b^{\gb_1} \ldots b^{\gb_t}\,,
\ee
where $a_\ga$ and $b^\gb$ are auxiliary oscillators
\be
\label{osc}
[a_{\ga}\,, a_{\gb}]_{} =0\z
[a_{\ga}\,, b^{\gb}]_{} =\delta^\gb_\ga \z
[b^{\ga}\,, b^{\gb}]_{} =0
\ee
being
generating elements of the star product algebra realization of the
generalized conformal higher spin symmetry. Note that the
expressions for the full set of generalized higher spin conformal
currents  (\ref{clo}) are much simpler than those for the
conformal higher spin currents in Minkowski spacetime \cite{KVZ}.

\section{Generalized Stress Tensors}
\label{Generalized Stress Tensors}
To present expression for
$T_{\ga_1 \ldots \ga_n}$ in terms  of the dynamical fields it is
useful to introduce a set of chains of
totally symmetric multisvector fields
$c_{\ga_1 \ldots \ga_n}^k$ of all ranks $n$
which satisfy the conditions
\be
\label{inter}
\f{\p}{\p X^{\beta_1\beta_2}}
c_{\ga_1 \ldots \ga_n}^k  (X)=
c_{\gb_1 \gb_2 \ga_1 \ldots \ga_n}^k (X)\,.
\ee
The Chan-Paton index $k$ enumerates different chains and can take an
arbitrary number of values. Let us set
\be
\label{Tphi}
T^{kl}_{\ga_1 \ldots \ga_n} (X) = \sum_{m=0}^n a((-1)^m , n)
\f{i^m\,n!}{m!(n-m)!} c_{\ga_1 \ldots \ga_m}^k (X)
c_{\ga_{m+1} \ldots \ga_n}^l (X)
\ee
with totally symmetrized indices $\ga$ on the right hand side and
arbitrary normalization coefficients $a((-1)^m, n)$.
The key fact is that the multisvector
$T^{kl}_{\ga_1 \ldots \ga_n} $ defined this way satisfies the
generalized conservation condition. To see this it is most
convenient to use its form (\ref{cond1}). It follows
\bee
&{}&\ls\ls\ls\lambda^{ \ga_1 \ldots \ga_n\,,\gb_1 \gb_2}
\f{\p}{\p X^{\gb_1 \gb_2}}
\Big (c_{ \ga_1 \ldots \ga_m} \Big )
c_{ \ga_1 \ldots \ga_{n-m}}
=\lambda^{ \ga_1 \ldots \ga_m \gga_1 \ldots\gga_{n-m}\,,\ga_{m+1} \ga_{m+2}}
c_{ \ga_1 \ldots \ga_{m+2}}
c_{ \gga_1 \ldots \gga_{n-m}}\nn\\
&{}&=\f{(n-m)(n-m-1)}{(m+2)(m+1)}
\lambda^{ \ga_1 \ldots \ga_{m+2} \gga_1 \ldots\gga_{n-m-2}\,,
\gga_{n-m-1} \gga_{n-m}}
c_{ \ga_1 \ldots \ga_{m+2}} c_{ \gga_1 \ldots
\gga_{n-m}}\nn\\
&{}&=\f{(n-m)(n-m-1)}{(m+2)(m+1)}
\lambda^{ \ga_1 \ldots \ga_{m+2} \gga_1 \ldots\gga_{n-m-2}\,,
\gb_1 \gb_2}
c_{ \ga_1 \ldots  \ga_{m+2}}
\f{\p}{\p X^{\gb_1 \gb_2}}c_{ \gga_1 \ldots\gga_{n-m-2}}\,,
\eee
where the property that symmetrization over any $n+1$ indices
in $\lambda^{ \ga_1 \ldots \ga_n \,,\gb_{1} \gb_{2}}$ gives zero
was used. As a result, because of a sign factor produced by the factor
of $i^m$ in (\ref{Tphi}) all terms in (\ref{cond1}) cancel pairwise.

The set of quantities $c_{\ga_1 \ldots \ga_n}^k$ satisfying
(\ref{inter}) is provided by the higher derivatives of the dynamical
scalar and svector fields
\be
\label{phb}
c_{\ga_1 \ldots \ga_{2p}}^k =
\f{\p}{\p X^{\ga_1 \ga_2}}
\f{\p}{\p X^{\ga_3 \ga_4}} \ldots
\f{\p}{\p X^{\ga_{2p-1} \ga_{2p}}} c^k(X)\,,
\ee
\be
\label{phf}
c_{\ga_1 \ldots \ga_{2p+1}}^k =
\f{\p}{\p X^{\ga_1 \ga_2}}
\f{\p}{\p X^{\ga_3 \ga_4}} \ldots
\f{\p}{\p X^{\ga_{2p-1} \ga_{2p}}} c_{\ga_{2p+1}}^k (X)\,.
\ee
Such defined
quantities $c_{\ga_1 \ldots \ga_n}^k$ are totally symmetric
in $\ga_1 \ldots \ga_n $ as a result
of the equations of motion (\ref{oscal}) and (\ref{ofer}) which imply that
any antisymmetrization of svector indices in higher derivatives of the
dynamical fields gives zero (note that every solution of (\ref{ofer})
satisfies (\ref{oscal}) \cite{Mar}).

In fact, the equation (\ref{inter})
is just the unfolded form of the equations (\ref{oscal}) and
(\ref{ofer}) from which they were derived in \cite{BHS}. In terms of the
generating function
\be
C^k(b|X)=\sum_{n=0}^\infty \f{1}{n!} c_{\ga_1 \ldots \ga_n}^k (X)
b^{\ga_1}\ldots b^{\ga_n}
\ee
the equation (\ref{inter}) reads
\be
\f{\p}{\p X^{\beta_1\beta_2}}C^k(b|X)=
\f{\p^2 }{\p b^{\beta_1} \p b^{\beta_2}}C^k(b|X) \,.
\ee
The generating function $C^k(b|X)$ was interpreted in \cite{BHS}
as a set of Fock modules
\be
|C^k(b|X)\rangle =C^k(b|X){} \vac
\ee
$(k= 1,2,3\ldots $) generated by the creation operators $b^\ga$ from
the vacuum state $\vac$ satisfying
\be
a_\ga {} \vac =0\z \vac {} b^\ga =0\,.
\ee

In terms of the analogous generating function for $T$
\be
T^{kl}(b|X)=\sum_{n=0}^\infty \f{1}{n!} T^{kl}_{\ga_1 \ldots \ga_n} (X)
b^{\ga_1}\ldots b^{\ga_n}
\ee
the formula (\ref{Tphi}) with $a((-1)^m, n) =1$ gets remarkably simple form
\be
T^{kl}(b|X)= C^k (ib |X) C^l (b|X) \,.
\ee
A proof of the conservation condition (\ref{cond1}) is now obvious
because,  in terms of the generating function
\be
\Lambda^{\gb_1 \gb_2}(a)=\sum_{n=0}^\infty \f{1}{n!}
\lambda^{\ga_1 \ldots \ga_n\,,\gb_1 \gb_2}  a_{\ga_1}\ldots a_{\ga_n}\,,
\ee
the Young property of $\lambda^{\ga_1 \ldots \ga_n\,,\gb_1 \gb_2}$
is equivalent to
\be
\label{propl}
\lambda^{ \ga_1 \ldots \ga_{n} \,,\gb_1 \gb_2}
a_{\ga_1}\ldots a_{\ga_n} a_{\gb_1} =0\,.
\ee
The generalized conservation property (\ref{cond1}) is equivalent to
\be
\label{prop}
\langle \Lambda^{\gb_1 \gb_2}(a) | {}
\f{\p}{\p X^{\gb_1 \gb_2}}
|T^{kl}(b|X)\rangle =0\,,
\ee
where $\langle \Lambda^{\gb_1 \gb_2}(a)| =\vac {}\Lambda^{\gb_1 \gb_2}(a) $
and $|T^{kl}(b|X)\rangle = T^{kl}(b|X)\vac$.
One gets
\bee
\langle \Lambda^{\gb_1 \gb_2}(a) | {}
\f{\p}{\p X^{\gb_1 \gb_2}}
|T^{kl}(b|X)\rangle &=&
\langle \Lambda^{\gb_1 \gb_2}(a) | {}\Big (
C^k (ib |X)\f{\p^2}{\p b^{\beta_1} \p b^{\beta_2}}( C^l (b|X))\nn\\
&-&\f{\p^2}{\p b^{\beta_1} \p b^{\beta_2}}( C^k (ib |X)) C^l (b|X)\Big )
{}\vac\,.
\eee
The property (\ref{propl}) implies that a total derivative
with respect to
$\f{\p}{\p b^{\gb_{1,2}}}$ does not contribute,
thus allowing to ``integrate by parts",
that immediately proves (\ref{prop}).

The formula (\ref{Tphi}) describes several different cases.
If $n$ is even (odd), the charges (\ref{charge}) have
commuting (anticommuting) parameters $\eta$. Let us define
generalized spin $s=\f{n}{2}$. Then
$T^{kl}_{\ga_1 \ldots \ga_{2s}}$
generate symmetries and supersymmetries for $s$ integer and half-integer,
respectively.

Using the ambiguity in the coefficients $a((-1)^m, n) $ in (\ref{Tphi})
one can fix statistics of the fields in
$T^{kl}_{\ga_1 \ldots \ga_{2s}}$, arriving at the following different cases
\be
\label{bb}
T_{BB}^{kl}{}_{\ga_1 \ldots \ga_{2s}} = \sum_{q=0}^s
\f{(-1)^q\,(2s)!}{(2q)!(2s-2q)!} c_{\ga_1 \ldots \ga_{2q}}^k c_{\ga_{2q+1}
\ldots \ga_{2s}}^l \qquad s\quad \mbox{integer}\,,
\ee
\be
\label{ff}
T_{FF}^{kl}{}_{\ga_1 \ldots \ga_{2s}} = \sum_{q=0}^{s-1}
\f{i(-1)^q\,(2s)!}{(2q+1)!(2s-2q-1)!} c_{\ga_1 \ldots \ga_{2q+1}}^k
c_{\ga_{2q+2} \ldots \ga_{2s}}^l \qquad s\quad \mbox{integer}\,,
\ee
\be
\label{bf}
T_{BF}^{kl}{}_{\ga_1 \ldots \ga_{2s}} = \sum_{q=0}^{s-\half}
\f{(-1)^q\,(2s)!}{(2q)!(2s-2q)!} c_{\ga_1 \ldots \ga_{2q}}^k c_{\ga_{2q+1}
\ldots \ga_{2s}}^l \qquad s\quad \mbox{half-integer} \,,
\ee
\be
\label{fb}
T_{FB}^{kl}{}_{\ga_1 \ldots \ga_{2s}} = \sum_{q=0}^{s-\half}
\f{i(-1)^q\,(2s)!}{(2q+1)!(2s-2q-1)!} c_{\ga_1 \ldots \ga_{2q+1}}^k
c_{\ga_{2q+2} \ldots \ga_{2s}}^l \qquad s\quad \mbox{half-integer}\,.
\ee
One observes that, for $s$ integer,  $T$ is symmetric
in the inner indices for even spins and antisymmetric for odd spins
\be
T^{kl}_{\ga_1 \ldots \ga_{2s}}=(-1)^s
T^{lk}_{\ga_1 \ldots \ga_{2s}}\,.
\ee
This formula holds both for the bosonic case of $T_{BB}$ and for
the fermionic  case of $T_{FF}$ provided that bosonic
and fermionic fields are, respectively, commuting and
anticommuting as required by microcausality \cite{Mar}.
These properties are in agreement with the standard symmetry properties
of usual spin 1 current, spin 2 stress tensor and their higher spin
generalizations. For the fermionic case one gets
\be
T_{BF}^{kl}{}_{\ga_1 \ldots \ga_{2s}}=i(-1)^{s+\half}
T_{FB}^{lk}{}_{\ga_1 \ldots \ga_{2s}}\,.
\ee

The following comment is now in order. According to (\ref{inter})
the components $c_{\ga_1 \ldots \ga_n}$ themselves satisfy the
conservation property (\ref{cond1}) and, therefore, can be used as
$T_{\ga_1 \ldots \ga_n}$ in the construction of currents.
By virtue of antisymmetrization
over $M+1$ indices, taking into account the field equations
(\ref{inter}), one can see that all forms containing
parameters $\eta_{\gb_1 \ldots \gb_t}{}^{\ga_1 \ldots \ga_s}$
with $s>0$ are equivalent modulo exact forms
to the analogous forms containing traces of $\eta$. This allows one
to consider only the case with $s=0$,
i.e. only the forms $\Omega (\eta )$ with the parameters
$\eta_{\gb_1 \ldots \gb_t}$ with $ t= 0,1,2,\ldots $ parametrize the
cohomology class associated with nontrivial integrals of
motion of this type. This result is expected because such parameters
are associated with the shifts
\be
\delta c(X) = \eta
+\eta_{\ga\gb} X^{\ga\gb}
+\eta_{\ga\gb\gga\eta} X^{\ga\gb}X^{\gga\eta}\ldots \z
\delta c_\ga (X) = \eta_{\ga}+\eta_{\ga\gb\gga} X^{\gga\gb}\ldots\,,
\ee
which are obvious symmetries of the equations (\ref{oscal}) and (\ref{ofer}).

\section{Fourier Transform and Invariant Norm}
\label{Fourier Transform and Invariant Norm}

The equations (\ref{oscal}) and (\ref{ofer}) were analyzed in
\cite{Mar} by means of Fourier transform. For a scalar field
\be
\label{expb}
c(X) = c_0 \exp i k_{\ga\gb}X^{\ga\gb}\,
\ee
 (\ref{oscal}) requires
\be
\label{ktw}
k_{\ga\gb} =\pm \xi_\ga \xi_\gb
\ee
with an arbitrary commuting real svector $\xi_\ga$.
The equation for a svector field $c_\ga (X)$
fixes in addition a polarization factor so that
the generic solution of the equations (\ref{oscal}) and (\ref{ofer})
has the form
\be
\label{bfo}
c (X) =c^+ (X) +c^- (X)\z
c^\pm (X)= \f{1}{\pi^{\f{M}{2}}}
\int d^M\xi\,  b^\pm (\xi ) \exp \pm i  \xi_{\ga}\xi_{\gb} X^{\ga\gb}
\,,
\ee
\be
\label{ffo}
c_\gga (X) =c_\gga^+ (X) +c_\gga^- (X)\z
c^\pm_\gga (X) =\f{1}{\pi^{\f{M}{2}}}
\int d^M\xi\, \xi_\gga
 f^\pm (\xi ) \exp \pm i  \xi_{\ga}\xi_{\gb} X^{\ga\gb}
\,.
\ee
The space of solutions is parametrized by two
functions of $M$ variables $\xi_\ga$
both for the scalar $c(X)$ and for the svector $c_\ga (X)$.
Scalar and svector therefore
have equal numbers of on-mass-shell degrees of freedom.
Because odd functions $b^\pm (\xi)$ and even functions
$f^\pm (\xi)$ do not contribute to (\ref{bfo})
and (\ref{ffo}), respectively, we demand
\be
 b^\pm (\xi)=b^\pm (-\xi) \,,\qquad
f^\pm (\xi)=-f^\pm (-\xi)\,.
\ee

In \cite{Mar} it was shown that charges generating
$osp(1,2M)$ symmetry and its higher spin extension
admit simple realization in terms
of Fourier components $b^\pm (\xi)$ and $f^\pm(\xi) $
\be
\label{TB}
Q^{B,F} (\eta ) =
\int d^M \xi   \Big (
b^+ (\xi ) \eta^{B,F} (\xi, \f{\p}{\p \xi} ) b^- (\xi ) +
f^+ (\xi ) \eta^{B,F} (\xi, \f{\p}{\p \xi} ) f^- (\xi ) \Big )\,.
\ee
Depending on the oddness properties, the parameters
\be
\label{par}
\eta (\xi, \f{\p}{\p \xi} ) = \sum_{kl}
\eta^{\ga_1 \ldots \ga_k}{}_{\gb_1 \ldots \gb_l}
\xi_{\ga_1}\ldots \xi_{\ga_k}
\f{\p}{\p \xi_{\gb_1}}\ldots \f{\p}{\p \xi_{\gb_l}}\,
\ee
are bosonic or fermionic
\be
\eta^B (-\xi, -\f{\p}{\p \xi} )=\eta^B (\xi, \f{\p}{\p \xi} )\z
\eta^F (-\xi, -\f{\p}{\p \xi} )=-\eta^F (\xi, \f{\p}{\p \xi} )\,.
\ee
Taking into account the
commutation relations introduced in \cite{Mar}
\be
\label{qpr}
[b^\pm (\xi_1 ) , b^\pm (\xi_2 )] =0\,,\qquad
[b^- (\xi_1 ) , b^+ (\xi_2 )] =
\half ( \delta (\xi_1 - \xi_2 )+\delta (\xi_1 + \xi_2 ) )\,,
\ee
\be
\label{qprf}
[f^\pm (\xi_1 ) , f^\pm (\xi_2 )]_+ =0\,,\qquad
[f^- (\xi_1 ) , f^+ (\xi_2 )]_+ =
\half ( \delta (\xi_1 - \xi_2 )- \delta (\xi_1 + \xi_2 ) )\,,
\ee
where $[,]_+$ denotes anticommutator, it is easy
to see that the charges (\ref{TB}) generate all
generalized higher spin transformations.

The analogy between
the symmetry parameters (\ref{par}) and the
parameters (\ref{eta}) in the closed form (\ref{clo}) suggests that
the expression for the generators (\ref{TB})
must result from the charges (\ref{charge}).
Let us show that this is indeed true.
Inserting (\ref{bfo}) and (\ref{ffo}) into (\ref{phb}) and (\ref{phf})
we get
\be
c^{\pm k}_{\ga_1 \ldots \ga_q} =(\pm i)^{[q/2]}
\f{1}{\pi^{\f{M}{2}}} \int d^M\xi\,  c^\pm (\xi ) \xi_{\ga_1}
\ldots \xi_{\ga_q} \exp \pm i  \xi_{\gga}\xi_{\gb} X^{\gga\gb}\,,
\ee
where $c^\pm (\xi ) =b^\pm (\xi ) $ for $q$ even
and $c^\pm (\xi ) =f^\pm (\xi ) $ for $q$ odd ($[r]$ denotes
the integer part of $r$).
Inserting this into (\ref{clo}), (\ref{Tphi}) with
\be
a ((-1)^m , n ) = i^{[\frac{n}{2}+\frac{1+(-1)^m}{4}]} \,,
\ee
we find for $\eta = 1$
\bee
Q(c^+ , c^- )  &=&
\epsilon_{\gamma_1 \ldots \gamma_M}\int_S
dX^{\gamma_1 \ga_1}\wedge \ldots \wedge dX^{\gamma_M \ga_M}
       \int d^M \xi  d^M \xi^\prime
c^+ (\xi ) c^- (\xi^\prime )\nn\\&{}& (\xi +\xi^\prime )_{\ga_1} \ldots
(\xi +\xi^\prime )_{\ga_M}
    \exp -iX^{\gga\gb} (\xi^\prime_\gga \xi^\prime_\gb -\xi_\gga \xi_\gb )\,.
\eee
Let an integration surface be a hyperplane parametrized by some coordinates
$y^i$ with $i=1\ldots M$, i.e. $X^{\ga\gb} = U^{\ga\gb}_i y^i$ where
the matrix $U^{\ga\gb}_i$ has rank $M$. It follows
\be
\label{Qin}
Q(c^+ , c^- )  =
 \int d^M y \int d^M \xi d^M \xi^\prime
det\Big | V^{\ga}_n (\xi + \xi^\prime)\Big|
\delta^M \Big ((\xi_\ga - \xi^\prime_\ga)
V^{\ga}_n (\xi + \xi^\prime)\Big )\,
c^+ (\xi ) c^- (\xi^\prime ) \,,
\ee
where
\be
V^{\ga}_n (\xi + \xi^\prime)=
 (\xi_\gb + \xi^\prime_\gb) U^{\ga\gb}_n\,.
\ee
The factor of $det\Big | V^{\ga}_n (\xi + \xi^\prime)\Big|$
guarantees that, generically,
 the zeros of the delta function in (\ref{Qin})
at $\xi_\gb + \xi^\prime_\gb \to 0$ do not contribute. As a result,
only the zeros $\xi_\gb - \xi^\prime_\gb \to 0$
have to be taken into account. The integration measure in
(\ref{Qin}) therefore amounts to  $\delta^M (\xi - \xi^\prime)$.
As a result one is left  with the expected expression
\bee
\label{norm}
Q(c^+ , c^- )  = \int d^M \xi c^+ (\xi ) c^- (\xi ) \,.
\eee
It remains to note  that, $Q(c^\pm , c^\pm ) =0$
because the bilinear form in  $\xi$ and $\xi^\prime$ in the exponential
$\exp \pm iX^{\gga\gb} (\xi^\prime_\gga \xi^\prime_\gb +\xi_\gga \xi_\gb )$
is semi-definite and, therefore, the argument of the delta-function
resulting from the integration over $y_n$ has no nontrivial zeros.

The integral of motion (\ref{norm}) produces an
invariant norm on the space of solutions.
It gives rise to a positive-definite invariant norm
\be
A({\bf b}^+, {\bf f}^+;{\bf b}^- ,{\bf f}^- )= \int d\xi^M \Big (
{\bf b}^+ (\xi )  {\bf b}^- (\xi ) + {\bf f}^+ (\xi )  {\bf f}^-
(\xi ) \Big )\,
\ee
on the one-particle quantum states
\be
 \int d^M \xi \Big ({\bf b}^- (\xi ) b^+ (\xi ) +
{\bf f}^- (\xi ) f^+ (\xi ) \Big )  \vac\,,
\ee
parametrized by the functions ${\bf b}^-$ and ${\bf f}^-$.

The charges with multispinor parameters
$\eta_{\gb_1 \ldots \gb_t}{}^{\ga_{1} \ldots \ga_{s}}$ give rise to
the higher spin charges (\ref{TB}) because every factor of $X^{\ga\gb}$
in (\ref{clo}) is equivalent to the second derivative over $\xi$.

\section{Conclusion}
\label{Conclusion}

It is shown that $sp(2M)$ invariant
equations of motion in the generalized spacetime
$\M_M$ with matrix coordinates suggested in \cite{BHS} 
admit conserved (super)charges associated with the
infinite-dimensional higher spin superextension of $osp(1,2M)$.
The (super)charges
are integrals of on-mass-shell closed $M$-forms bilinear in the
dynamical fields. This provides one more manifestation of the fact
that nontrivial independent degrees of freedom
of the $sp(2M)$ invariant dynamics live on $M$-dimensional surfaces
\cite{Mar}. The scalar charge gives rise to
 an invariant norm on the space of solutions of the
field equations which turns out to be equivalent to the positive-definite 
     norm on the Fock  space of one-particle states.
The proposed construction has a good chance to admit a
generalization to less trivial (not necessarily flat) geometries.

It is straightforward to write down conformally invariant Noether 
current interactions as
\be
\label{No}
S^{Noether} = \int_{ST} \Omega(A) \,,
\ee
where $\Omega(A)$ is the $M+1$ form obtained from the
$M$-form (\ref{clo}) via replacement of the parameter
$\eta_{\ga_1 \ldots \ga_n}{}^{\gb_1 \ldots \gb_m}$
by the higher spin conformal gauge 1-form
$A_{\ga_1 \ldots \ga_n}{}^{\gb_1 \ldots \gb_m}$. 
The integration in (\ref{No}) is performed over a $M+1-$dimensional 
surface to be interpreted as spacetime \cite{Mar}.

The dynamics described by (\ref{oscal}) and (\ref{ofer}) is equivalent 
\cite{BHS,Mar} to the conformal dynamics in Minkowski spacetime for specific
(infinite for $M>2$) sets of usual relativistic fields contained in
scalar and svector fields in $\M_M$.
Conserved charges built in this paper amount to certain sets
of conserved charges in the associated Minkowski spacetime.
In fact they give rise to infinite towers of different higher spin charges
built of various conformal fields in the usual Minkowski spacetime.
(Recall that infinite towers of fields in Minkowski space result from
``Kaluza-Klein" modes on the fiber of the local Cauchy fiber bundle.)
An interesting problem is to analyze the content
of the conserved higher spin
currents associated with the set of symmetry parameters
$\eta_{\gb_1 \ldots \gb_t}{}^{\ga_{1} \ldots \ga_{s}}$
from the perspective of Minkowski spacetime. It should be
taken into account that the conformal Minkowski fields hidden in $c(X)$ and
$c_\ga (X)$ contain higher derivatives when expressed in terms of
gauge potentials.
For example, $c(X)$ describes scalar field in the spin 0
sector, Maxwell field strength in the spin 1 sector, Weyl tensor in
the spin 2 sector etc. Generically, a spin $s$ field contains
order $[s]$ derivatives of the respective conformal gauge potential. This
is in agreement with the fact that, for example, the generalized
conformal stress tensor in the spin two sector is built from the Weyl
tensor, thus corresponding to the stress tensor of the $C^2$ conformal
(Weyl) gravity.

Technically, to derive explicit form of all conformal higher spin
currents built from various pairs of Minkowskian conformal fields
is going to be a hard problem which requires a knowledge of a content of
the decomposition of a totally symmetric tensor product of an arbitrary
number of spinor representations into irreducible representations of
the Lorentz algebra. To the best of our knowledge
a solution of this problem is yet unknown for generic $M$. This problem 
can be avoided however in case one manages to operate in totally
$sp(2^p)$ invariant terms of $sp(2^p)$
multiplets or rather $osp(N,2^p)$ supermultiplets.

\section*{Acknowledgments}
This research was supported in part by INTAS, Grant No.00-1-254,
the RFBR Grant No.02-02-17067 and the RFBR Grant No.01-02-30024.

\end{document}